\documentstyle[twoside]{article}

\catcode`\@=11
\long\def\@makefntext#1{
\protect\noindent \hbox to 3.2pt {\hskip-.9pt  
$^{{\eightrm\@thefnmark}}$\hfil}#1\hfill}               

\def\@makefnmark{\hbox to 0pt{$^{\@thefnmark}$\hss}}    
        
\def\ps@myheadings{\let\@mkboth\@gobbletwo
\def\@oddhead{\hbox{}
\rightmark\hfil\eightrm\thepage}   
\def\@oddfoot{}\def\@evenhead{\eightrm\thepage\hfil
\leftmark\hbox{}}\def\@evenfoot{}
\def\sectionmark##1{}\def\subsectionmark##1{}}

\oddsidemargin=\evensidemargin
\newcounter{sectionc}\newcounter{subsectionc}\newcounter{subsubsectionc}
\renewcommand{\section}[1] {\vspace{12pt}\addtocounter{sectionc}{1} 
\setcounter{subsectionc}{0}\setcounter{subsubsectionc}{0}\noindent 
        {\tenbf\thesectionc. #1}\par\vspace{5pt}}
\renewcommand{\subsection}[1] {\vspace{12pt}\addtocounter{subsectionc}{1} 
        \setcounter{subsubsectionc}{0}\noindent 
        {\bf\thesectionc.\thesubsectionc. {\kern1pt \bfit #1}}\par\vspace{5pt}}
\renewcommand{\subsubsection}[1] {\vspace{12pt}\addtocounter{subsubsectionc}{1}
        \noindent{\tenrm\thesectionc.\thesubsectionc.\thesubsubsectionc.
        {\kern1pt \tenit #1}}\par\vspace{5pt}}
\newcommand{\nonumsection}[1] {\vspace{12pt}\noindent{\tenbf #1}
        \par\vspace{5pt}}

\newcounter{appendixc}
\newcounter{subappendixc}[appendixc]
\newcounter{subsubappendixc}[subappendixc]
\renewcommand{\thesubappendixc}{\Alph{appendixc}.\arabic{subappendixc}}
\renewcommand{\thesubsubappendixc}
        {\Alph{appendixc}.\arabic{subappendixc}.\arabic{subsubappendixc}}

\renewcommand{\appendix}[1] {\vspace{12pt}
        \refstepcounter{appendixc}
        \setcounter{figure}{0}
        \setcounter{table}{0}
        \setcounter{lemma}{0}
        \setcounter{theorem}{0}
        \setcounter{corollary}{0}
        \setcounter{definition}{0}
        \setcounter{equation}{0}
        \renewcommand{\thefigure}{\Alph{appendixc}.\arabic{figure}}
        \renewcommand{\thetable}{\Alph{appendixc}.\arabic{table}}
        \renewcommand{\theappendixc}{\Alph{appendixc}}
        \renewcommand{\thelemma}{\Alph{appendixc}.\arabic{lemma}}
        \renewcommand{\thetheorem}{\Alph{appendixc}.\arabic{theorem}}
        \renewcommand{\thedefinition}{\Alph{appendixc}.\arabic{definition}}
        \renewcommand{\thecorollary}{\Alph{appendixc}.\arabic{corollary}}
        \renewcommand{\theequation}{\Alph{appendixc}.\arabic{equation}}
        \noindent{\tenbf Appendix \theappendixc #1}\par\vspace{5pt}}
\newcommand{\subappendix}[1] {\vspace{12pt}
        \refstepcounter{subappendixc}
        \noindent{\bf Appendix \thesubappendixc. {\kern1pt \bfit #1}}
        \par\vspace{5pt}}
\newcommand{\subsubappendix}[1] {\vspace{12pt}
        \refstepcounter{subsubappendixc}
        \noindent{\rm Appendix \thesubsubappendixc. {\kern1pt \tenit #1}}
        \par\vspace{5pt}}

\newcommand{\textlineskip}{\baselineskip=13pt}
\newcommand{\smalllineskip}{\baselineskip=10pt}

\def\eightcirc{
\begin{picture}(0,0)
\put(4.4,1.8){\circle{6.5}}
\end{picture}}
\def\eightcopyright{\eightcirc\kern2.7pt\hbox{\eightrm c}} 

\newcommand{\copyrightheading}[1]
        {\vspace*{-2.5cm}\smalllineskip{\flushleft
        {\footnotesize International Journal of Modern Physics C, #1}\\
        {\footnotesize $\eightcopyright$\,\,\, World Scientific Publishing
         Company}\\
         }}

\def\abstracts#1#2#3{{
        \centering{\begin{minipage}{4.5in}\baselineskip=10pt\footnotesize
        \parindent=0pt #1\par
        \parindent=15pt #2\par
        \parindent=15pt #3\par
        \end{minipage}}\par}}

\newcommand{\bibit}{\nineit}
\newcommand{\bibbf}{\ninebf}
\renewenvironment{thebibliography}[1]
        {\frenchspacing
         \ninerm\baselineskip=11pt
         \begin{list}{\arabic{enumi}.}
        {\usecounter{enumi}\setlength{\parsep}{0pt}     
         \setlength{\leftmargin 17pt}{\rightmargin 0pt}   
         \setlength{\itemsep}{0pt} \settowidth
        {\labelwidth}{#1.}\sloppy}}{\end{list}}

\newcounter{itemlistc}
\newcounter{romanlistc}
\newcounter{alphlistc}
\newcounter{arabiclistc}

\newcommand{\fcaption}[1]{
        \refstepcounter{figure}
        \setbox\@tempboxa = \hbox{\footnotesize Fig.~\thefigure. #1}
        \ifdim \wd\@tempboxa > 5in
           {\begin{center}
        \parbox{5in}{\footnotesize\smalllineskip Fig.~\thefigure. #1}
            \end{center}}
        \else
             {\begin{center}
             {\footnotesize Fig.~\thefigure. #1}
              \end{center}}
        \fi}

\newcommand{\tcaption}[1]{
        \refstepcounter{table}
        \setbox\@tempboxa = \hbox{\footnotesize Table~\thetable. #1}
        \ifdim \wd\@tempboxa > 5in
           {\begin{center}
         \parbox{5in}{\footnotesize\smalllineskip Table~\thetable. #1}
            \end{center}}
        \else
             {\begin{center}
             {\footnotesize Table~\thetable. #1}
              \end{center}}
        \fi}

\def\@citex[#1]#2{\if@filesw\immediate\write\@auxout
        {\string\citation{#2}}\fi
\def\@citea{}\@cite{\@for\@citeb:=#2\do
        {\@citea\def\@citea{,}\@ifundefined
        {b@\@citeb}{{\bf ?}\@warning
        {Citation `\@citeb' on page \thepage \space undefined}}
        {\csname b@\@citeb\endcsname}}}{#1}}

\newif\if@cghi
\def\cite{\@cghitrue\@ifnextchar [{\@tempswatrue
        \@citex}{\@tempswafalse\@citex[]}}
\def\citelow{\@cghifalse\@ifnextchar [{\@tempswatrue
        \@citex}{\@tempswafalse\@citex[]}}
\def\@cite#1#2{{$\null^{#1}$\if@tempswa\typeout
        {IJCGA warning: optional citation argument 
        ignored: `#2'} \fi}}

\def\pmb#1{\setbox0=\hbox{#1}
        \kern-.025em\copy0\kern-\wd0
        \kern.05em\copy0\kern-\wd0
        \kern-.025em\raise.0433em\box0}

\def\fnt#1#2{\footnotetext{\kern-.3em
        {$^{\mbox{\scriptsize #1}}$}{#2}}}

\def\fpage#1{\begingroup
\voffset=.3in
\thispagestyle{empty}\begin{table}[b]\centerline{\footnotesize #1}
        \end{table}\endgroup}

\def\runninghead#1#2{\pagestyle{myheadings}
\markboth{{\protect\footnotesize\it{\quad #1}}\hfill}
{\hfill{\protect\footnotesize\it{#2\quad}}}}
\font\tenbf=cmbx10
\font\tenit=cmti10 
\font\tenit=cmti10
\font\bfit=cmbxti10 at 10pt
\font\ninebf=cmbx9
\font\ninerm=cmr9
\font\nineit=cmti9

\font\eightrm=cmr8

\def\lsym{\raise-3pt\hbox{\vbox{\tabskip0pt\offinterlineskip
        \halign{\tabskip0pt plus 1em
        ##\tabskip0pt\cr
        $\,\,<\,\,$\cr
        $\,\,\sim\,\,$\cr}}}}
\def\rsym{\raise-3pt\hbox{\vbox{\tabskip0pt\offinterlineskip
     \halign{\tabskip0pt plus 1em
      ##\tabskip0pt\cr
      $\,\,>\,\,$\cr
      $\,\,\sim\,\,$\cr}}}}
\def\qed{\hbox{${\vcenter{\vbox{                        
        \hrule height 0.4pt\hbox{\vrule width 0.4pt height 6pt
        \kern5pt\vrule width 0.4pt}\hrule height 0.4pt}}}$}}
\def\theequation{\thesection.\arabic{equation}}         

\def \ind#1{{\mbox{\scriptsize {#1}}}}
\input{epsf}
\begin{document}

\runninghead{K. Vollmayr-Lee, W. Kob, K. Binder \& A. Zippelius}
{Cooling Rate Dependence and Dynamic Heterogeneity}

\normalsize\textlineskip
\thispagestyle{empty}
\setcounter{page}{1}

\copyrightheading{Vol. 0, No. 0 (2000) 000--000}

\vspace*{0.88truein}

\fpage{1}
\centerline{\bf COOLING RATE DEPENDENCE AND DYNAMIC HETEROGENEITY}
\vspace*{0.035truein}
\centerline{\bf BELOW THE GLASS TRANSITION IN A LENNARD-JONES GLASS} 
\vspace*{0.37truein}
\centerline{\footnotesize K. Vollmayr-Lee$^1$, W. Kob$^2$, K. Binder$^2$ and 
A. Zippelius$^3$} 
\vspace*{0.015truein}
\centerline{\footnotesize\it $^1$Department of Physics,
Bucknell University, } 
\baselineskip=10pt
\centerline{\footnotesize\it Lewisburg, PA 17837, USA}
\vspace*{0.15truein}
\centerline{\footnotesize and}
\vspace*{0.15truein}
\centerline{\footnotesize\it $^2$Institute of Physics,
Johannes-Gutenberg-University Mainz}
\vspace*{0.015truein}
\centerline{\footnotesize\it Staudinger Weg 7, 55099 Mainz, Germany}
\vspace*{0.15truein}
\centerline{\footnotesize and}
\vspace*{0.15truein}
\centerline{\footnotesize\it $^3$Institute of Theoretical Physics,
Georg-August-University G\"ottingen}
\vspace*{0.015truein}
\centerline{\footnotesize\it Bunsenstr.9, 37073 G\"ottingen, Germany}

\vspace*{0.225truein}

\vspace*{0.21truein}
\abstracts{
We investigate a binary Lennard-Jones mixture with molecular dynamics 
simulations. We consider first a system cooled linearly in time
with the cooling rate $\gamma$. By varying 
$\gamma$ over almost four decades we study the influence of the cooling 
rate on the glass transition and on the resulting glass. We find for all
investigated quantities a cooling rate dependence; with decreasing
cooling rate the system falls out of equilibrium at decreasing
temperatures, reaches lower enthalpies and obtains increasing local
order. Next we study the dynamics of the melting process
by investigating the most immobile and most mobile particles in the glass. 
We find
that their spatial distribution is heterogeneous and that the immobile/mobile
particles are surrounded by denser/less dense cages than an average particle.
}{}{}



\vspace*{1pt}\textlineskip      
\section{Introduction}          
\vspace*{-0.5pt}
\noindent
If a liquid is cooled rapidly enough it avoids crystallization and falls
out of equilibrium. The resulting system is in an amorphous state and
shows dramatically different dynamics than the high temperature 
liquid.\cite{ediger_1996} Computer simulations provide access to 
microscopic information of these interesting static and dynamic 
features. 

In this paper we will present molecular dynamics simulations of a binary
Lennard-Jones mixture which has been shown to be not prone to 
crystallization.\cite{kob_andersen} Our goal is to investigate both 
the influence of the cooling rate on the statics of the glass, as well as the
dynamics of melting as the glass is heated.

The outline of this paper is as follows. In the next section we present
the interaction model and details of the simulation. In section 3 we
investigate the cooling rate dependence, 
first by studying the influence of the cooling
rate on the cooling process, i.e. the glass transition, and second, by
investigating the resulting end configurations and the
microscopic origin of the cooling rate dependence. More
extensive reports can be found elsewhere\cite{LJ_work}, as well as similar 
cooling rate studies for amorphous silica\cite{SiO2_work}. 
In section 4 we investigate the dynamics of a melting
process via the spatial distribution and the surrounding of the 
fastest and the slowest particles {\em below} the glass transition  
(similar to the work of Kob {\it et al.}\cite{NIST_work}
{\em above} the glass transition). We finish with a summary.

\setcounter{section}{2}
\setcounter{equation}{0}
\section{Model and Details of the Simulation}
\noindent We use a binary Lennard-Jones mixture where the two particle types, 
A and B, have the same mass $m$. The interaction potential for particles
$i$ and $j$ at positions $\vec{r}_i$ and $\vec{r}_j$ and of type 
$\alpha,\beta \in \{$A,B$\}$ is
\begin{equation}  
V_{\alpha \beta}(r) = 4 \epsilon_{\alpha \beta} 
\left ( \left ( \frac{\sigma_{\alpha \beta}}{r} \right )^{12}   
      - \left ( \frac{\sigma_{\alpha \beta}}{r} \right )^{6}
\right ),
\end{equation}
where $r=|\vec{r}_i-\vec{r}_j|$. 
To avoid crystallization the potential parameters are chosen to be 
$\epsilon_\ind{AA}=1.0 , \epsilon_\ind{AB}=1.5 , \epsilon_\ind{BB}=0.5 , 
\sigma_\ind{AA}=1.0 , \sigma_\ind{AB}=0.8$ and 
$\sigma_\ind{BB}=0.88$.\cite{kob_andersen}  We 
use reduced units: the unit of length is $\sigma_\ind{AA}$, the 
unit of energy is $\epsilon_\ind{AA}$ and the unit of time is 
$\sqrt{m \sigma_\ind{AA}^2/48 \epsilon_\ind{AA}}$. 
For the molecular dynamics simulations we use the velocity Verlet
algorithm with a step size $\Delta t=0.02 \tau$. All simulations are done
with 800 A and 200 B particles. To get better statistics all quantities are 
averaged over 10 configurations.

For section 3 we use the constant pressure algorithm proposed by
Andersen\cite{const_p} with the piston mass $M=0.05$ and the external 
pressure $p_\ind{ext}=0$. For section 4 the microcanonical ensemble is 
used.

\bigskip
\setcounter{section}{3}
\setcounter{equation}{0}
\section{Cooling Rate Dependence}
\vspace*{-0.5pt}
\noindent
In this section we investigate how the glass transition and the
resulting glass depend on its history. In specific, we cool the system
from a high temperature, $T_0=2.0$, to zero temperature by coupling it to
a stochastic heat bath. The bath temperature $T_\ind{b}$ is
decreased linearly in time $t$ with the cooling rate $\gamma$, i.e. 
\begin{equation}  
T_\ind{b}(t)= T_0 - \gamma \, t .
\end{equation}
We simulated 13 different cooling rates, ranging from $\gamma=0.02$
to $\gamma = 3.125 \cdot 10^{-6}$. 

\subsection{Glass Transition}
\noindent
Here we investigate the system during the cooling process. 
In Fig.~1 we show the enthalpy $H=U_\ind{pot} + U_\ind{kin} +
\frac{M}{2} \dot{V}^2 + p_\ind{ext} V$ as a function of the heat bath
temperature $T_\ind{b}$.
\begin{figure}




  \epsfxsize=7.2cm
  \centerline{
  \epsfbox{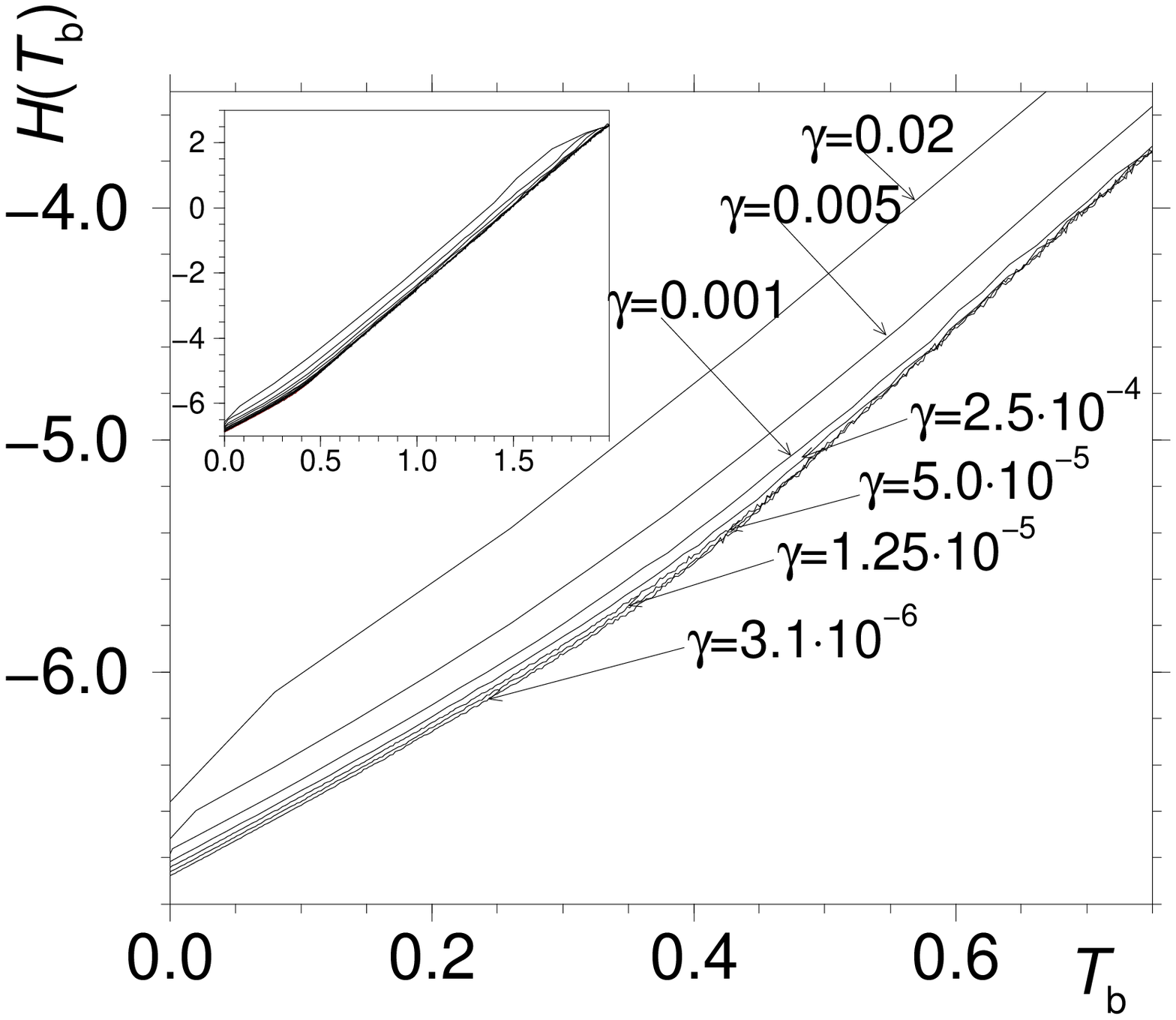}
              } 
\fcaption{Enthalpy $H$ as a function of the heat bath temperature
$T_\ind{b}$. For clarity we show only every second cooling rate.}
\end{figure}
At high temperatures the curves are the same for all cooling rates
$\gamma$; the system is still in equilibrium. At lower bath
temperatures the system falls out of equilibrium and bends off the
equilibrium curve. This bending is often associated with the 
onset of the glass transition and occurs when the system relaxation time is 
approximately the inverse of the cooling rate.
Therefore, faster cooling rates result in a higher temperature at
which the system falls out of equilibrium.

To specify the temperature at which the system falls out of equilibrium
we use the fictive temperature $T_\ind{g}$ proposed by Tool 
and Eichlin\cite{tool}.
\begin{figure}




  \epsfxsize=7.6cm
  \centerline{
  \epsfbox{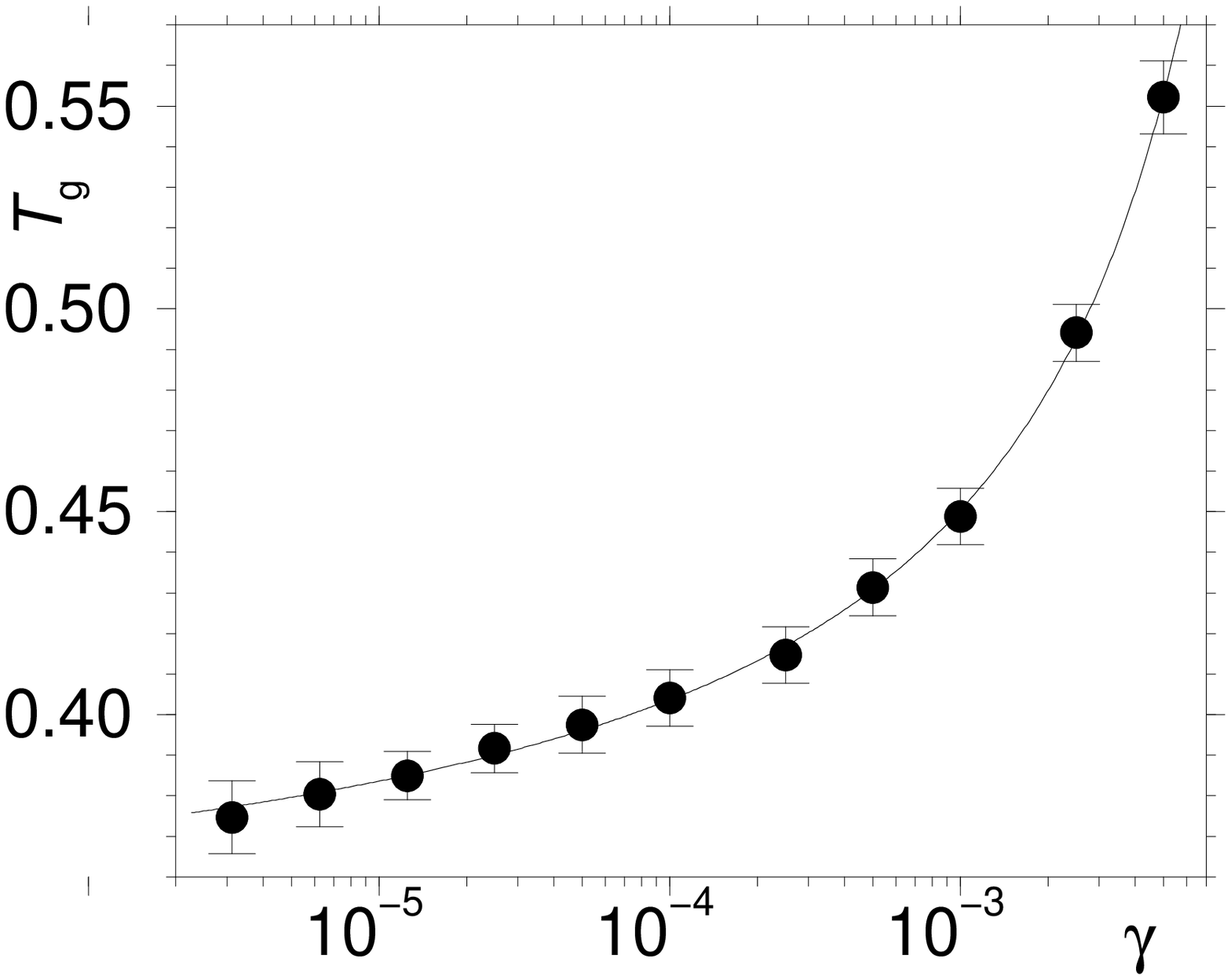}
              } 
\fcaption{Fictive Temperature $T_\ind{g}$ as a function of $\gamma$.
Included is a Vogel-Fulcher fit.}
\end{figure}
This is defined as the temperature at which the high and low $T_\ind{b}$
extrapolations of $H(T_\ind{b})$ intersect.
Fig.~2 shows the cooling rate dependence of $T_\ind{g}$, where, as expected,
$T_\ind{g}$ decreases with
decreasing cooling rate $\gamma$. Also included in the figure is a
Vogel-Fulcher fit $T_\ind{g}=T_0+\frac{B}{\ln{(A \gamma)}}$, which provides 
with $T_0=0.348$, $A=31.820$ and $B=0.403$ a good fit to our data.

\subsection{Resulting Glass at Zero Temperature}
\noindent
We now investigate the system after it has been cooled to 
zero temperature, i.e.\ the end configurations.
\begin{figure}




  \epsfxsize=7.6cm
  \centerline{
  \epsfbox{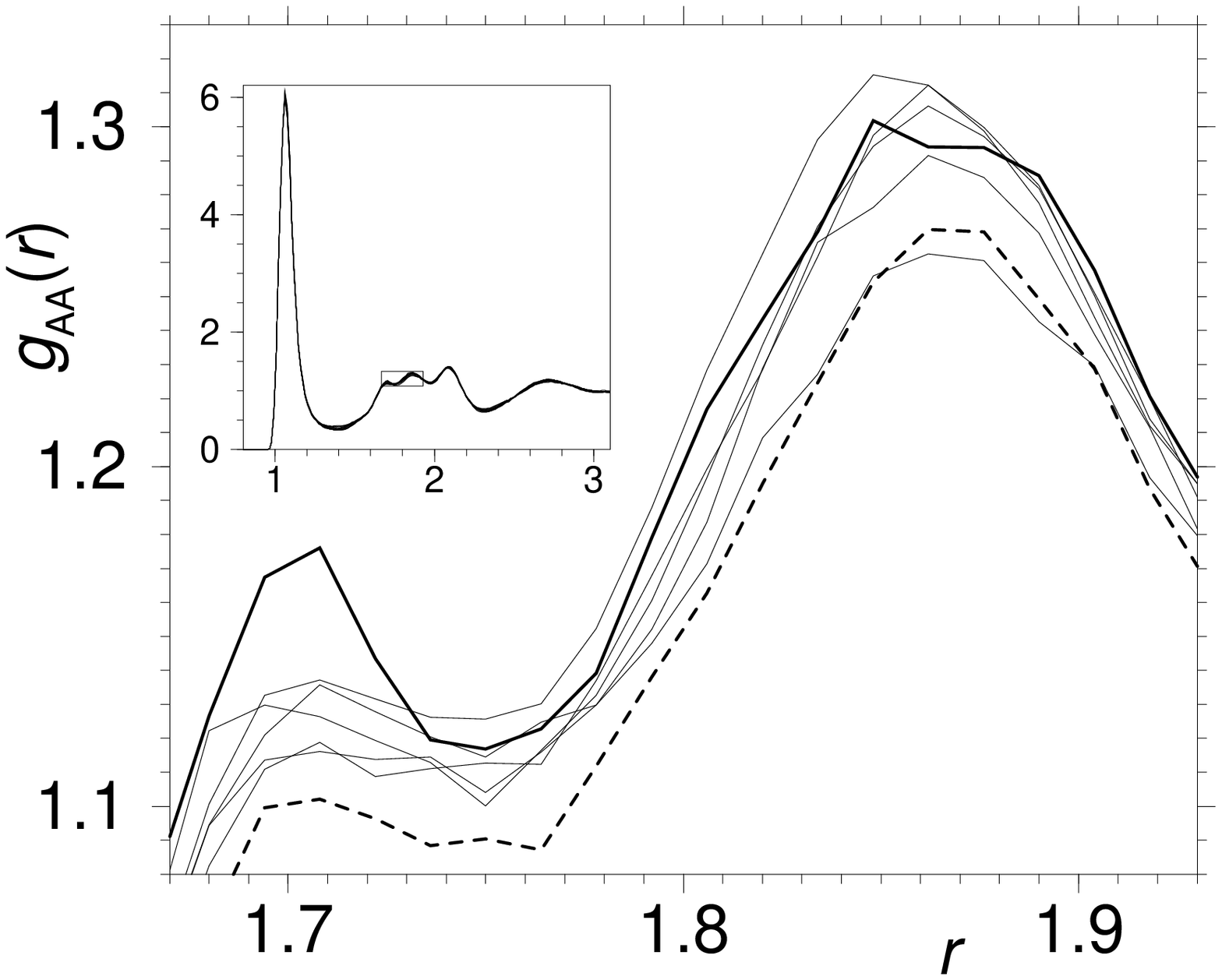}
              } 
\fcaption{Radial pair distribution for A-particles. 
The slowest/fastest cooling rate corresponds to the solid/dashed bold line.
The figure shows an enlargement of the second nearest neighbor 
peak as indicated with a box in the inset.}
\end{figure}
Fig.~3 shows the radial distribution function of the A-particles. We find
with decreasing cooling rate more pronounced peaks. The same effect
is found in the bond-bond angle distribution (see
Fig.~4). For the latter we used that two particles are defined to be
neighbors and connected by a bond if their distance is less than the
position of the first minimum of the corresponding (by particle type) 
$g(r)$. 
\hspace*{5mm}
\begin{figure}




  \epsfxsize=7.3cm
  \centerline{
  \epsfbox{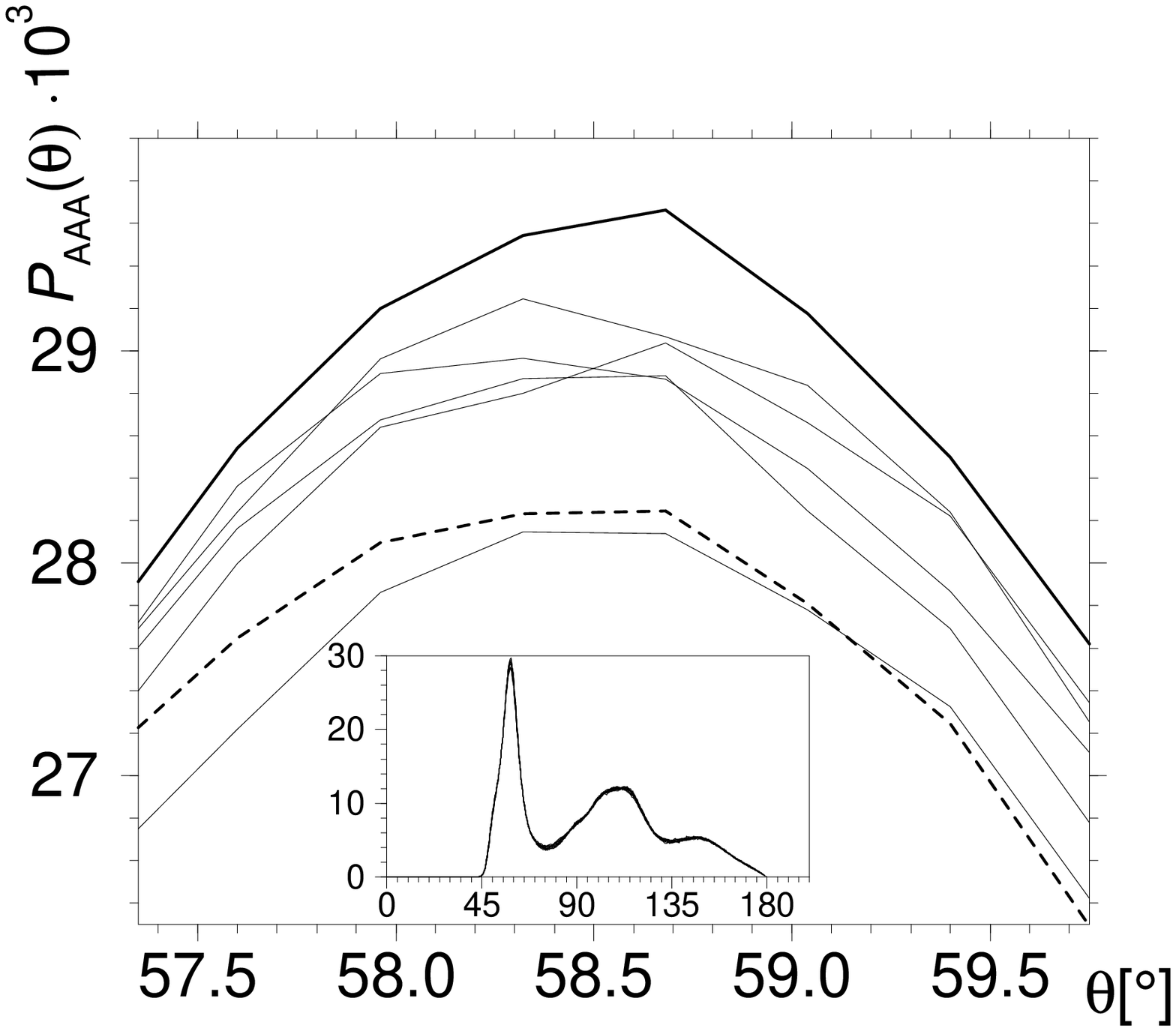}
              } 
\fcaption{Bond-bond angle distribution for A-particles. 
The slowest/fastest cooling rate corresponds to the solid/dashed bold line.
The figure shows an enlargement of the first peak of the inset.}
\end{figure}
We find both for the radial distribution function and for the bond-bond angle 
distribution increasing order with decreasing cooling rate. 

In the previous subsection we have seen that the system reaches 
a lower enthalpy with decreasing cooling rate. We now address 
{\em how} the system lowers its energy with decreasing $\gamma$. The
question we ask is whether the system lowers its energy primarily by
rearranging the particles and their neighbors or instead by the change in 
distances and bond-bond angles, as shown in Fig.~3 and Fig.~4. To answer
this question we define a cluster as the set of a particle 
and its neighbors. We define a cluster type 
$\alpha_{\mu \nu}$ where the central particle is of type $\alpha \in
\{$A,B$\}$ with $\mu$ nearest neighbors of which $\nu$ are of
type B. The energy $E_\ind{cl}$ of a cluster is given by the sum of all
pairwise interactions between any two members of the cluster.
\begin{figure}




  \epsfxsize=7.3cm
  \centerline{
  \epsfbox{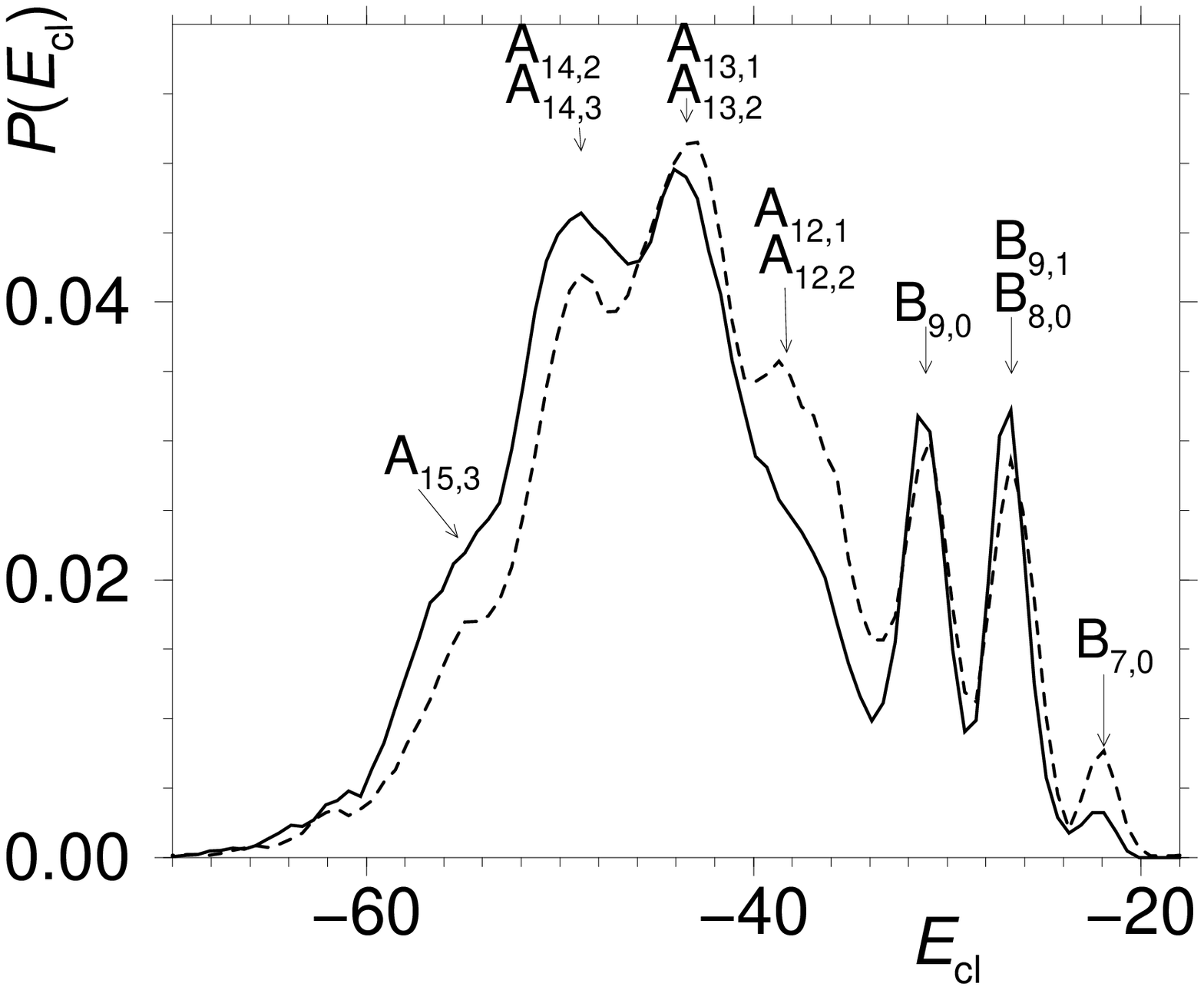}
              } 
\fcaption{Distribution of cluster energies as defined in the text. 
The slowest/fastest cooling rate corresponds to the solid/dashed bold line.}
\end{figure}
Fig.~5 shows the distribution of cluster energies and assigns the
different cluster types to the corresponding peak positions. The
distributions of the fastest and slowest cooling rate differ mainly 
in their peak height and only slightly in their peak position.
We therefore can conclude that
the energy is lowered primarily by a rearrangement of the particles.

\newpage
\bigskip
\setcounter{section}{4}
\setcounter{equation}{0}
\section{Melting: Dynamics Below The Glass Transition}
\vspace*{-0.5pt}
\noindent
Similar to the previous section we will again investigate the glass
after it has fallen out of equilibrium. So far we studied static
properties and in this section we will now investigate the dynamic properties
of the glass below the glass transition temperature $T_c=0.435$.
This $T_c$ corresponds to the glass transition temperature 
of mode coupling 
theory\cite{kob_andersen}. Instead of cooling the system we are now
asking the question how a glass melts. 

Our initial configurations are obtained via a rapid quench to $T=0.15$,
starting from a well equilibrated temperature at T=0.466.
We then continued in several steps. For each temperature we first equilibrated
with $10^5$ MD steps with a (NVT) simulation and then ran the production
run of $10^5$ MD steps with a (NVE) simulation. 
The volume is at all temperatures $V=831$.
Each equilibration and production run was followed by an instantaneous 
quench to the next higher temperature, where the process is 
repeated.

Having in mind the picture of an amorphous solid, we quantify the
dynamics of a particle by its average fluctuations from its 
average position, 
\begin{equation}  
d_i^2 = \overline{ \,\Big|\vec{r}_i(t)- 
                              \overline{ \vec{r}_i(t) } \,\Big|^2  } \qquad ,
\end{equation}
where the bar denotes an average over the whole time 
of the production run.
\begin{figure}




  \epsfxsize=7.2cm
  \centerline{
  \epsfbox{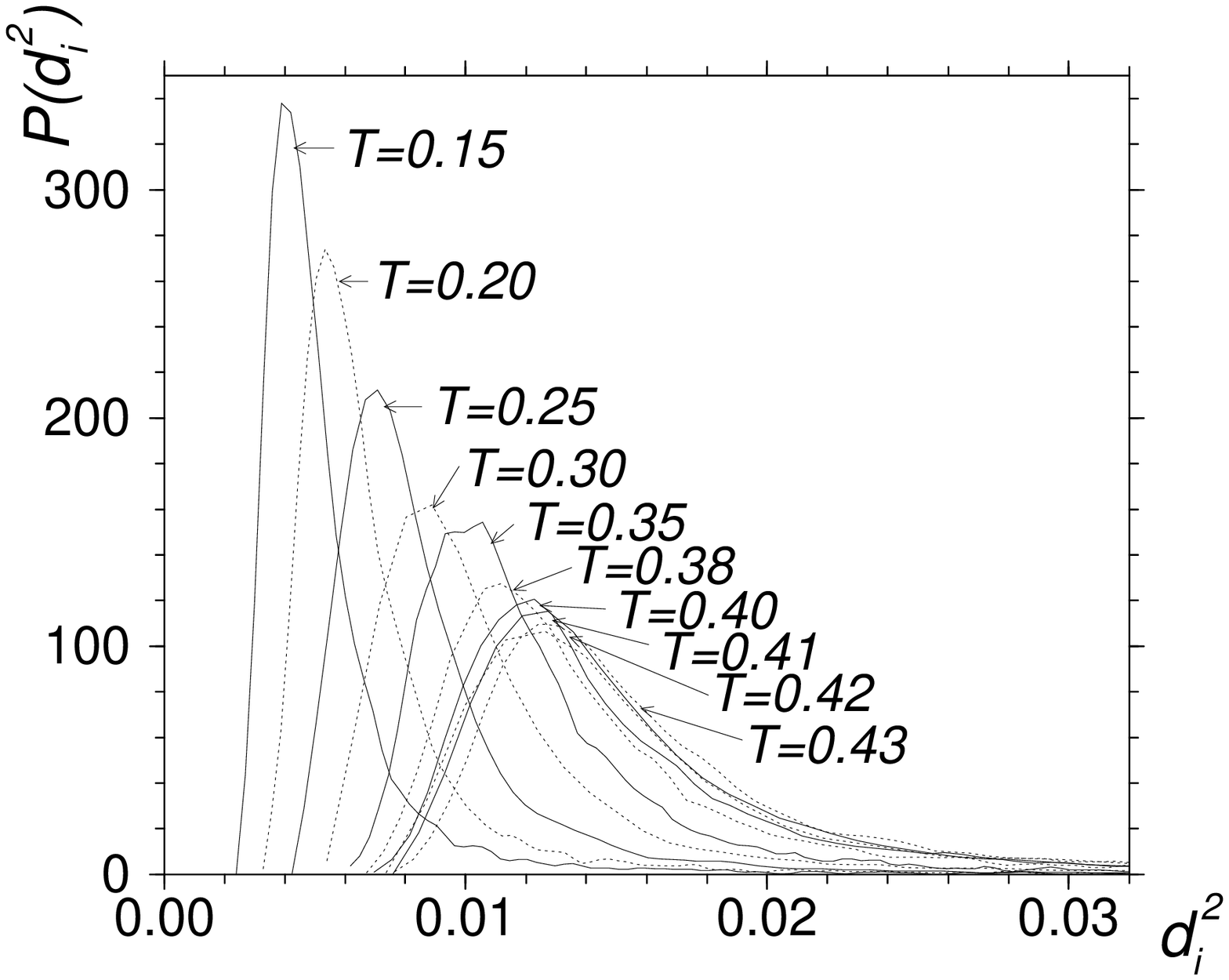}
              } 
\fcaption{$P(d_i^2)$ as defined in the text. The curves of 
all investigated temperatures are shown.}
\end{figure}
Fig.~6 shows the distribution of $d_i^2$ of all particles
from 10 independent runs and for all investigated temperatures. 
Most of the particles fluctuate around
their average site with $d_i^2 \approx 0.01$, i.e. they are frozen
into their positions and do not escape their cage of nearest neighbors
during the whole production run. The larger the temperature the more is
$P(d_i^2)$ shifted to the right.

We are interested in the fastest and slowest particles,
hence we define mobile/immobile particles  as the 5\%
with the largest/smallest $d_i^2$. 
We study the spatial distribution of these particles and also 
look at their surrounding in order to understand why they are more 
mobile or immobile.

\subsection{Spatial distribution of Mobile and Immobile Particles}
\noindent
\begin{figure}




  \epsfxsize=7.2cm
  \centerline{
   \epsfbox{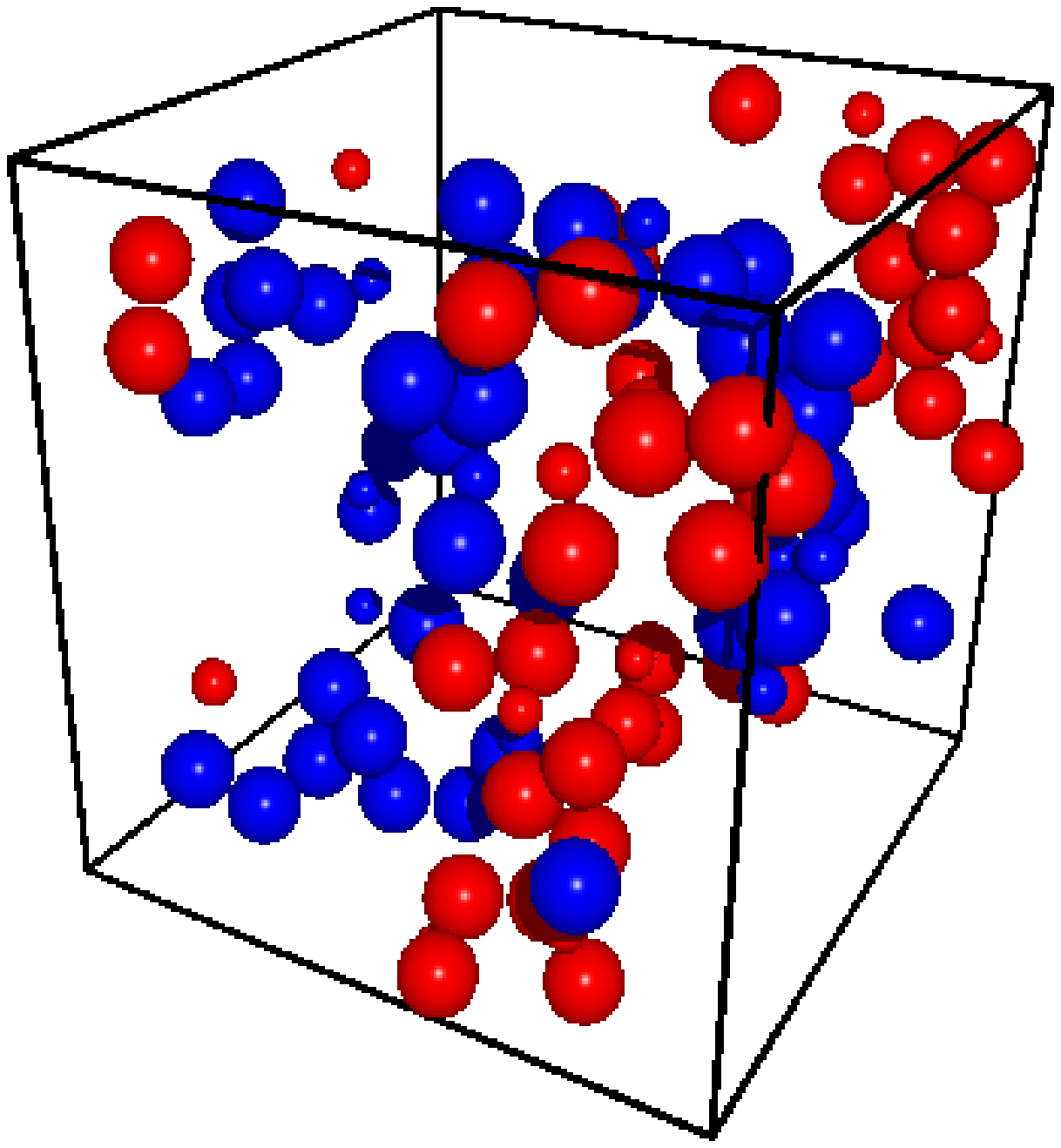}
              } 
\fcaption{Snapshot at $T=0.15$ 
of the mobile A particles (large \& light) and B particles
(small \& light) and the immobile A particles (large \& dark) and B
particles (small \& dark).}
\end{figure}
Fig.~7 shows a snapshot of the mobile and the immobile particles at 
$T=0.15$ and time $t=0$.  We find similar snapshots at all other times and
temperatures. The snapshot shows a clear heterogeneity. To quantify this
heterogeneity we use, similar to Ref. [5],  
the ratio of the radial distribution function of mobile A-particles, 
$g_\ind{mAmA}$, over that of all A particles, $g_\ind{AA}$ (see Fig. 8).
\begin{figure}




  \epsfxsize=7.2cm
  \centerline{
  \epsfbox{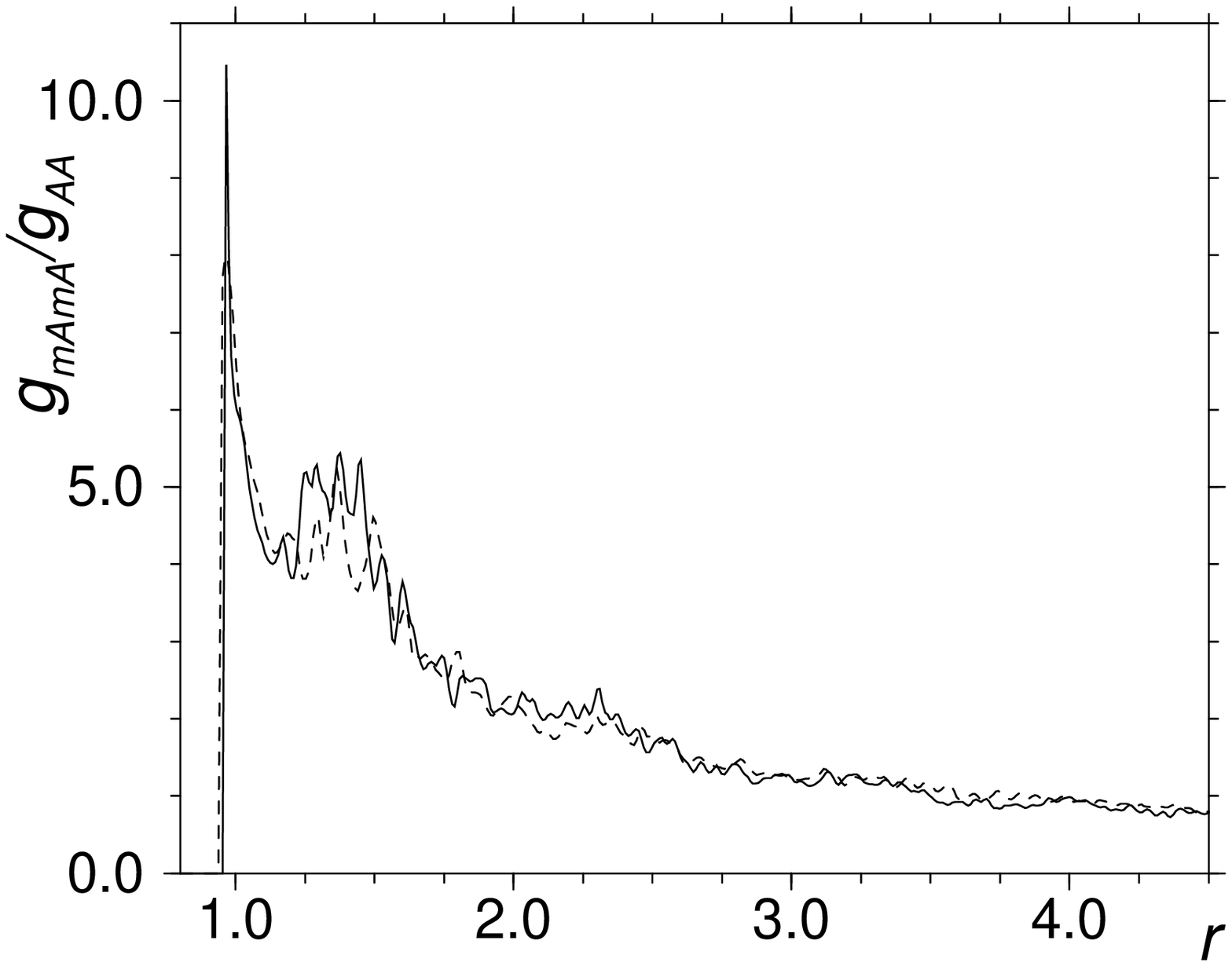}
              } 
\fcaption{$g_\ind{mAmA}/g_\ind{AA}$ as explained in the text. The 
solid line corresponds to $T=0.15$, the dashed line to $T=0.43$.}
\end{figure}
We find similar results for all temperatures and also for the
immobile particles. 
That the ratio $g_\ind{mAmA}/g_\ind{AA}$ differs significantly 
from unity indicates a strong dynamic heterogeneity.

\subsection{Neighborhood of Mobile and Immobile Particles}
\noindent
In this subsection we will investigate what causes a particle to be
mobile or immobile. 
\begin{figure}




  \epsfxsize=6.5cm
  \centerline{
  \epsfbox{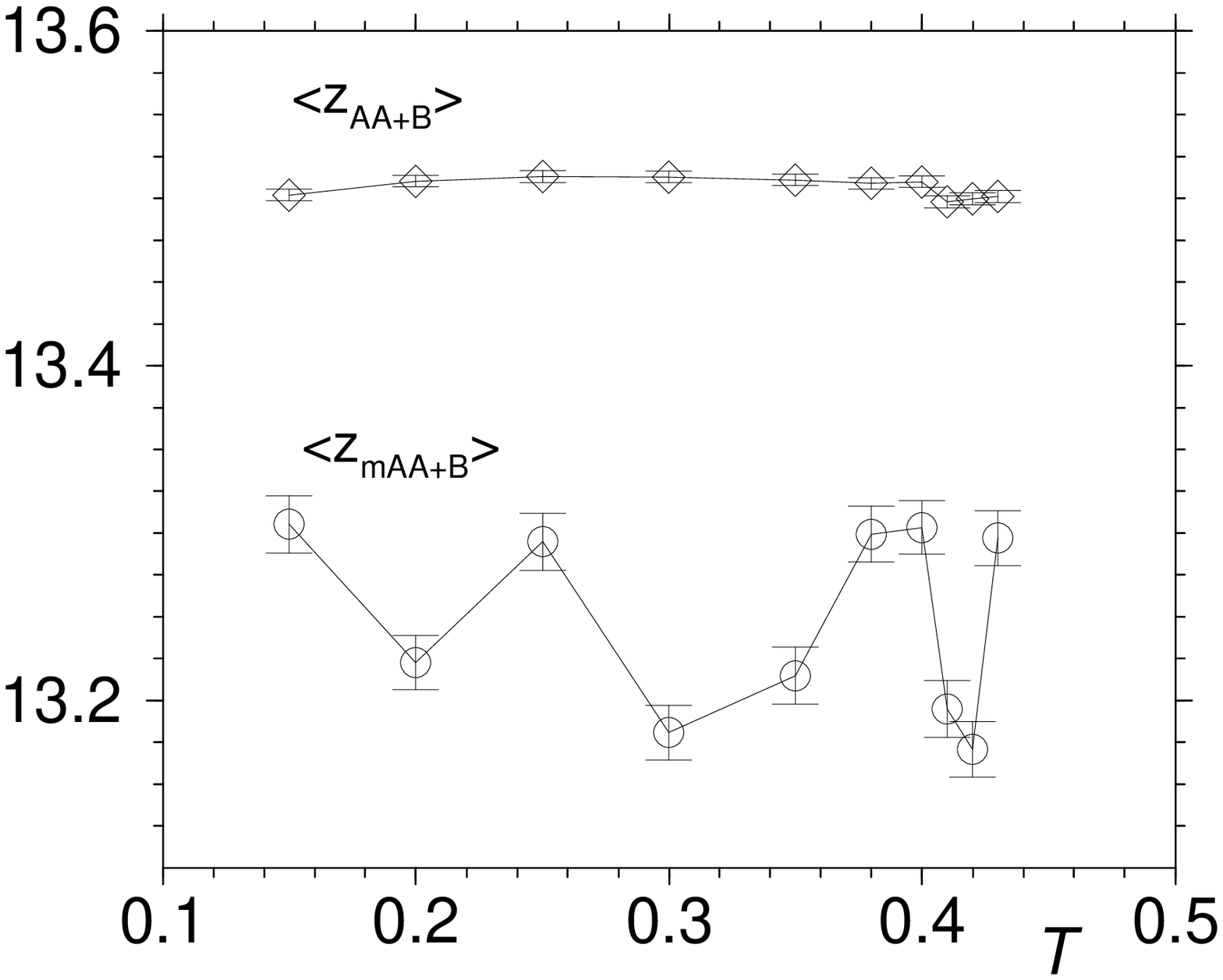}
              } 
\fcaption{Average total number of neighbors of a mobile A-particle 
($\langle z_\ind{mAA+B} \rangle$) in comparison with the total 
number of neighbors of any A-particle ($\langle z_\ind{AA+B} \rangle$).
The error bars correspond to the standard deviation of the mean. The real error bars are larger due to the lack of independent configurations over which the 
average has been taken.}
\end{figure}
Fig.~9 shows a comparison of the average total number of neighbors of a
mobile A-particle $\langle z_\ind{mAA+B} \rangle$ and the average 
total number of all A-particles, $\langle z_\ind{AA+B} \rangle$. 
We find that the mobile particle is
on average surrounded by fewer particles than usual.
\begin{figure}




  \epsfxsize=6.5cm
  \centerline{
  \epsfbox{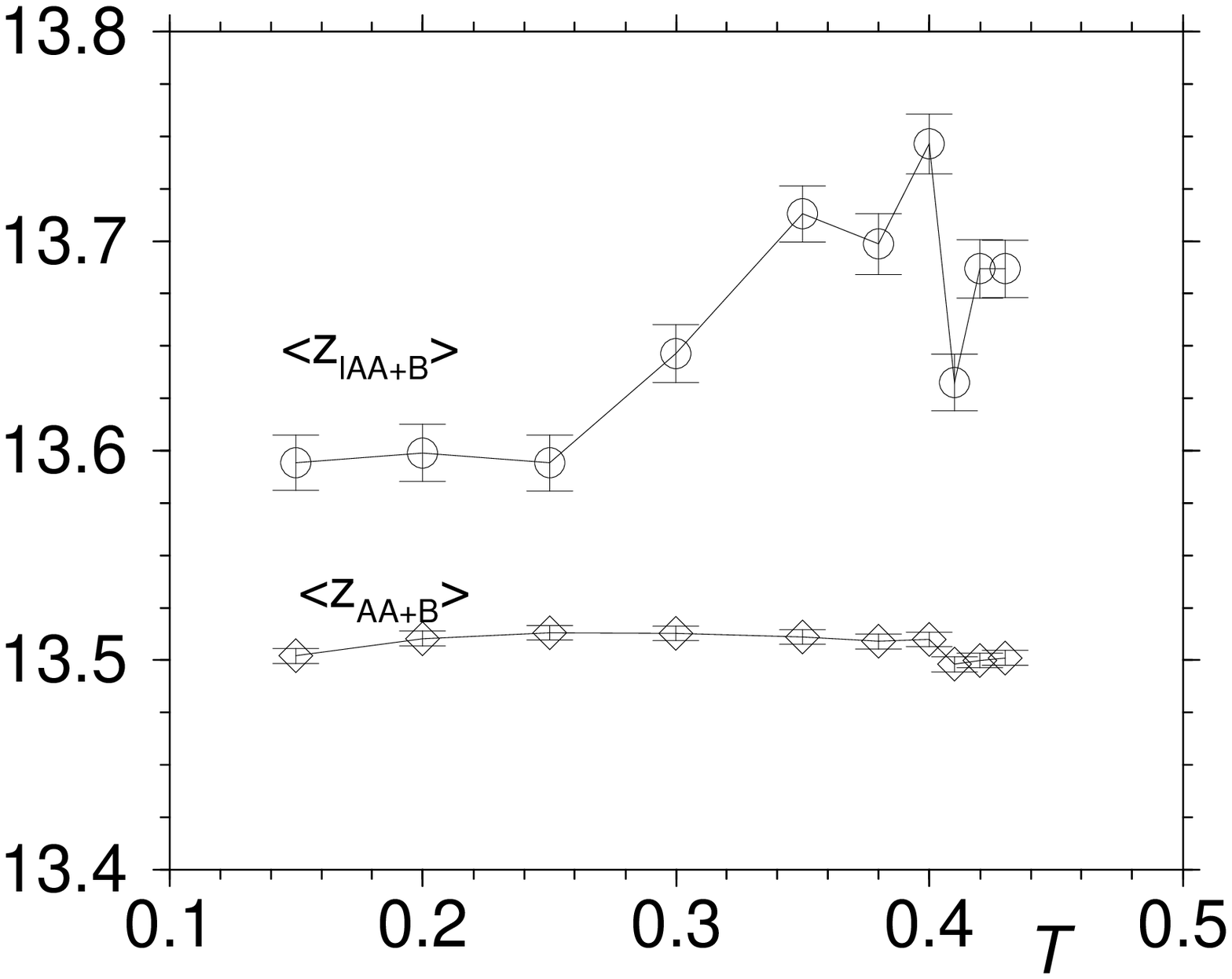}
              } 
\fcaption{Average total number of neighbors of an immobile A-particle, 
$\langle z_\ind{lAA+B} \rangle$, in comparison with the total 
number of neighbors of any A-particle, $\langle z_\ind{AA+B} \rangle$.}
\end{figure}
Fig.~10 shows conversely that an immobile (localized) 
A-particle with average total number of
neighbors $\langle z_\ind{lAA+B} \rangle$ has more particles
than usual surrounding it. The cage of mobile particles is less dense, the
cage of an immobile particle is denser.
\newpage
\section{Conclusion}
\noindent

We investigated both the static properties of the configurations
during and after the cooling to $T=0$ and the dynamical properties upon 
the subsequent heating of a binary Lennard-Jones mixture.
When cooling to $T=0$ we find for all investigated
quantities a cooling rate dependence; decreasing the cooling rate results in
an increasing order and a decreasing enthalpy. The energetic part of
the enthalpy is mainly lowered by a rearrangement of the particles. The
fictive temperature is decreasing with decreasing cooling rate and is
well fit by a Vogel-Fulcher law. 

During the melting we find dynamic heterogeneity;  the mobile and
immobile particles are strongly clustered. 
The motion of particles appears to be determined by its neighborhood, 
as the average coordination number is decreased for mobile particles,  
indicating a looser cage, and increased for immobile particles.

\nonumsection{Acknowledgment}
\noindent
K.V.-L. gratefully acknowledges financial support from
Schott-Glaswerke, Mainz, and the SFB 262.

\nonumsection{References}

\end{document}